\documentclass[aps,twocolumn,prb,showpacs,superscriptaddress]{revtex4}

\usepackage{graphicx}

\begin{document}

\title{The interplay between double exchange, super-exchange,\\ and
Lifshitz localization in doped manganites}

\author{J.~L.~Alonso}
\affiliation{Departamento de F{\'\i}sica Te\'orica, Facultad de Ciencias,
Universidad de Zaragoza, 50009 Zaragoza, Spain.}
\author{L.~A.~Fern\'andez} 
\affiliation{Departamento de F{\'\i}sica Te\'orica, 
Facultad de Ciencias F{\'\i}sicas,
Universidad Complutense de Madrid, 28040 Madrid, Spain.}
\author{F.~Guinea}
\affiliation{Instituto de Ciencia de Materiales (CSIC). Cantoblanco,
28049 Madrid. Spain.}
\author{V.~Laliena}
\affiliation{Departamento de F{\'\i}sica Te\'orica, Facultad de Ciencias, 
Universidad de Zaragoza, 50009 Zaragoza, Spain.}
\author{V.~Mart\'{\i}n-Mayor} \affiliation{Dipartimento di Fisica,
Universit\`a di Roma ``La Sapienza'', INFN, SMC and UdR1 of INFM,
P.le Aldo Moro 2, 00185 Roma, Italy.}
\date{\today}
\begin{abstract}
Considering the disorder caused in manganites by the substitution
Mn$\rightarrow$ Fe or Ga, we accomplish a systematic study of doped
manganites begun in previous papers. To this end, a disordered model
is formulated and solved using the Variational Mean Field
technique. The subtle interplay between double exchange,
super-exchange, and disorder causes similar effects on the dependence
of $T_\mathrm{C}$ on the percentage of Mn substitution in the cases
considered. Yet, in La$_{2/3}$Ca$_{1/3}$Mn$_{1-y}$Ga$_y$O$_3$ our
results suggest a quantum critical point (QCP) for $y\approx 0.1-0.2$,
associated to the localization of the electronic states of the
conduction band.  In the case of La$_x$Ca$_x$Mn$_{1-y}$Fe$_y$O$_3$
(with $x=1/3,3/8$) no such QCP is expected.
\end{abstract}
\maketitle

\section{Introduction}

A continuous phase transition at zero temperature defines a Quantum
Critical Point (QCP), with a diverging length-scale for
quantum-fluctuations and a diverging time scale for
dynamics.\cite{HERTZ76,SACHDEV,BELITZ-KIRK01}  The Renormalization
Group~\cite{WILSONKOGUT74} can be extended to the study of
QCP,\cite{HERTZ76} which has become a popular concept among
theoreticians, given the enhanced predictive power of the
calculations. For instance, the experimental finding of a QCP has been
postulated to underlie the physics of high-temperature
superconductors~\cite{CASTELLANI-DICASTRO} and
spin-ladders,\cite{DAGOTTO-RICE96,SACHDEV} it has been used to
predict the scaling behavior for meta-magnetic quantum criticality in
metals,\cite{MILLIS-etal01} or to study the effects of disorder in
quantum spin-chains.\cite{IGLOI01} The localization transition of non
interacting electrons in the presence of disorder~\cite{LEE-RAMA85} is
another type of continuous phase transition at zero temperature.  The
experimental finding of a QCP compatible with the theoretical
expectations for the Anderson metal-insulator
transition~\cite{LEE-RAMA85} would be very interesting. Note that the
Anderson transition is not the cause of the metal-insulator transition
in La$_{1-x}$Ca$_x$MnO$_3$,\cite{SMOLYANINOVA00} a well known
colossal magnetoresitance (CMR) manganite.\cite{CVM99}

In this paper, we want to address the behavior of  typical
CMR manganites when doped at the Mn sites
\begin{equation}
\mathrm{La}_{1-x}\mathrm{AE}_x\mathrm{Mn}_{1-y}\mathrm{TR}_y\mathrm{O}_3\, ,
\label{dobleserie}
\end{equation}
where AE\,=\,Ca,\,Sr and TR\,=\,Fe,\,Al,\,Ga, in the range
$0.3<x<0.5$.  Some experimental work has already been done for this
materials~\cite{SUN99,Blasco97,Ahn96,Sun98,Xavier00} but, from the
present analysis we believe that several interesting features have not
still been found experimentally. Specifically, we will show that, from
the knowledge gained in the study of the phase diagrams of
(La$_{1-z}$RE$_z$)$_{1-x}$Sr$_x$MnO$_3$ and
(La$_{1-z}$RE$_z$)$_{1-x}$Ca$_x$MnO$_3$, where RE is a trivalent rare
earth, and from our model calculations, one can conclude that:

i) When the $y=0$ material shows a first order transition from the
paramagnetic state (PM) to the ferromagnetic (FM) one, like in
La$_{2/3}$Ca$_{1/3}$MnO$_3$~\cite{FIRSTORDER}, a QCP should appear
close to 15\% Ga substitution ($x=1/3$ and $y=0.15$).  Unfortunately
previous experimental work~\cite{SUN99} did not go beyond 10\% of Ga,
but more experiments will be done.\cite{Algarabel02} In this case, it is
interesting that the localization transition determines the magnetic
behavior and then, through the corresponding {\em magnetic} critical
exponents, it may be possible to investigate the physical properties
of such QCP experimentally. In this context, note that disorder added
to a system with a low $T$ first order transition separating ordered
phases (see Sec.~\ref{sec:results} and Fig.~\ref{FIG2}) produces
quantum-critical like features.\cite{Burgy01}

ii) When the $y=0$ material shows a second-order PM---FM phase
transition, as in La$_{2/3}$Sr$_{1/3}$MnO$_3$, the low temperature
phase will be ferromagnetic until values of $y$ as large as $\approx
0.4$, and (if Mn is substituted with Ga) a FMmetal---FMinsulator
transition at finite temperature should take place (see also
Ref.~\onlinecite{Verges01}).

The layout of the rest of this paper is as follows.  In the next
section, we describe the experimentally known effects of Mn
substitution and we will discuss the simplest model that can account
for it. We will first consider the relatively simpler case of Fe
$\rightarrow$ Mn substitution, and we will then address the case of
Gallium. From now on, we will not be concerned with the Al case,
because this substitution causes the lattice to loose oxygen for
$y\approx 0.1$, and a significant amount of oxygen vacancies are
present.\cite{Blasco97} This phenomenon can easily be taken into
account with a proper modification of the present model, but this is
left for future work.  In section~\ref{sec:results} we present our
results, while the conclusions, comments, and predictions are left for
the last section.

\section{Model \label{sec:model}}

\subsection{Fe substitution}

In previous works,\cite{Alonso00a,Alonso00b,Alonso00c,Alonso01a}
we have considered the simplest
model that describes the magnetic properties of 
(La$_{1-z}$RE$_z$)$_{1-x}$AE$_x$MnO$_3$ and we concluded that, for
$0.3<x<0.5$ such model is the single orbital double exchange model
(DEM) on a cubic lattice \cite{DEM} with isotropic hopping and
super-exchange antiferromagnetic first-neighbors interaction:
\begin{equation}
{\cal H} =\sum_{ij} 
 t( \mbox{\boldmath$S$}_i,\mbox{\boldmath$S$}_j )
c_i^\dag c_j +
\sum_{\langle ij\rangle} {J}_\mathrm{AF} \mbox{\boldmath$S$}_i\cdot\mbox{\boldmath$S$}_j
\, ,\label{hamil}
\end{equation}
where $c_i$ corresponds to a Mn $\mathbf{e_g}$ orbital while
$\mbox{\boldmath$S$}$ stands for a unit vector oriented parallel to
the Mn$^{3+}$ ($S = 3/2$) core spins, which we assume to be
classical.\cite{classical}  The function $t(
\mbox{\boldmath$S$}_i,\mbox{\boldmath$S$}_j )=
t[\cos\frac{\theta_i}{2}\cos\frac{\theta_j}{2}+
\sin\frac{\theta_i}{2}\sin\frac{\theta_j}{2} {\mathrm e}^{{\mathrm
i}(\varphi_i-\varphi_j)}]$ stands for the overlap of two spin 1/2
spinors oriented along the directions defined by
$\mbox{\boldmath$S$}_i$ and $\mbox{\boldmath$S$}_j$, whose polar and
azimuthal angles are denoted by $\theta$ and $\varphi$, respectively.
The hopping integral, $t$, is approximately 0.16 eV.\cite{CVM99}  As
far as the magnetic interactions are concerned, one can trade the
effects of phonons,\cite{phonons} by tuning the superexchange
constant $J_\mathrm{AF}$.\cite{Alonso00a} The phase diagram of the
model~(\ref{hamil}) has been extensively studied by means of the
variational mean-field (VMF) technique~\cite{Alonso00a,Alonso00b} and
by the Hybrid Monte Carlo method.\cite{Alonso00c,Alonso01a} These
studies have shown that the parameter $J_\mathrm{AF}$ is rather
constrained. On one hand, if one wants to reproduce the first-order
character of the PM-FM phase
transition,\cite{FIRSTORDER} one should have $J_\mathrm{AF}$ larger
than $0.06t$.\cite{Alonso00a} But, on the other hand, the system is
FM at low temperatures only up to $J_\mathrm{AF}\approx 0.08t$, for a
hole concentration of 1/3.\cite{Alonso01a} The conclusion is twofold:
i) $J_\mathrm{AF}$ is not a very tunable parameter and, ii)
La$_{2/3}$Ca$_{1/3}$MnO$_3$ is a rather critical system where small
perturbations can produce large changes in physical properties.

Therefore, it is a challenge to this model to give account of the
experimental situation of the double series (\ref{dobleserie}) in the
range $0.3<x<0.5$ \cite{SUN99,Blasco97,Ahn96,Sun98,Xavier00}, without
stretching out the values of $J_{\mathrm{AF}}$ already
determined.\cite{Alonso00a,Alonso00b,Alonso01a}

Through the whole series (\ref{dobleserie}), only the Mn $\mathbf{e_g}$ (up)
band is electronically active, with the electron hopping between
Mn$^{3+}$ and Mn$^{4+}$. In the Fe case, the Fe $\mathbf{e_g}$ (up) band is
completely filled and electron hopping from Mn$^{3+}$ to Fe$^{3+}$ is
forbidden. The situation is similar when TR\,=\,Al$^{3+}$ or Ga$^{3+}$, in
which cases the d-band is full. This means that the fraction of
Mn$^{4+}$ with respect to the amount of Mn$^{3+}$+Mn$^{4+}$ is
incresed by a factor $(1-y)^{-1}$ respect to the case $y=0$. Figure 2
of Ref.~\onlinecite{Alonso00a} shows that, as far as the the following
argument is concerned, the Curie temperature, $T_{\mathrm C}$ does not depend
much on the hole density around 40\% of carriers. This
will allow us to extrapolate the results for TR\,=\,Fe from our previous
computations.  In fact, the similar ionic radii of Fe$^{3+}$ and
Mn$^{3+}$ means that lattice distortion effects may be ignored. Thus,
let us explore the hypothesis that the substitution of Mn for Fe
affects only the antiferromagnetic interaction between the localized
spins, besides the irrelevant in this case $(1-y)^{-1}$ factor in the effective
number of holes. This change is due to the arising of Mn-Mn, Mn-Fe,
and Fe-Fe couplings, with probability $(1-y)^2$, $2y(1-y)$ and $y^2$
respectively. Given that the core spins are 3/2 for Mn and 5/2 for Fe,
the effective value of the super-exchange constant is
\begin{equation}
J_{\mathrm{eff}}\;=\;J_{\mathrm{AF}}
\left[(1-y)^2+\frac{5}{3}2y(1-y)+\frac{25}{9}y^2\right]\, ,
\label{jeff}
\end{equation}
Thus, we can easily extrapolate the results of Fig. 3 of
Ref.~\onlinecite{Alonso00a} for $x=3/8$ to the case of Fe substitution
(we took $J_{\mathrm{AF}}/t\approx 0.07$ \cite{Alonso00a}), where we
get $J_{\mathrm{eff}}=1.17J_{\mathrm{AF}}$ for $y=0.12$, what implies
that $T_{\mathrm C}(y=0.12)/T_{\mathrm C}(y=0)=0.4$ (see Fig.~3 of
Ref.~\onlinecite{Alonso00a}), in complete agreement both with the
experimental data of Ref.~\onlinecite{Ahn96} as well as with the
phenomenological analysis of A. Tzavellas {\em et al.}.
\cite{Tzavellas00}

On the other hand, when TR in (\ref{dobleserie}) is Al or Ga, lattice
distortion effects enhance the relevance of the fact that on the Mn
sites the charge is a mixture of +3 and +4. In fact, as we will see,
electrostatic effects yielding diagonal disorder in our model turn out
to be crucial not only to understand the known experimental 
facts,\cite{SUN99,Blasco97,Algarabel02} but also to produce a new and very 
interesting physical situation for values of $y$ not studied in
Ref.~\onlinecite{SUN99}.

\subsection{Ga substitution}

The effects of Ga doping in the colossal magnetoresistance materials
La$_{2/3}$Ca$_{1/3}$MnO$_3$, has been recently studied by preparing
the series La$_{2/3}$Ca$_{1/3}$Mn$_{1-y}$Ga$_y$O$_3$
($y=0,0.02,0.05,0.1$),\cite{SUN99} and a strong decrease of $T_{\mathrm C}$
with growing $y$ was reported. However, the doping fraction never went
beyond 10\%, while we believe that a QCP will appear for $y \approx
0.1$---$0.2$. No structural change upon doping was
observed,\cite{SUN99} nor it is to be expected for larger Ga
fraction, because it does not occur in the LaGa$_y$Mn$_{1-y}$O$_3$
series. The model used here does not take into account the orbital
degeneracy in the $\mathbf{e_g}$ band. Thus, strictly speaking it ceases
to be valid when the ratio Mn$^{4+}$/Mn$^{3+}$ approaches unity,
which, in this case, corresponds to $y \approx 1/3$. Nevertheless we
use the results for $y$ up to 0.4 in order to describe better the
change in physical properties due to the transition.
 
The Ga substitution {\em disorders} the system (details are given
below).  In order to describe specifically a disorder effect, we will
need a {\em disordered model}.\cite{PARISI} The standard way of
constructing a disordered model is to choose first a simple model able
to describe the ordered situation (i.e. the system with no Gallium),
and then to mimic the inclusion of impurities in the real system by
the random modification of some terms in the Hamiltonian that reflect
the microscopic effects of the impurity. As said in the previous
subsection our reference {\em ordered} model, is the single orbital
Double-Exchange Model (DEM) on the cubic lattice of Eq.~(\ref{hamil}).

With an ordered model in our hands, we can discuss the effects of
Gallium doping. Given that the sizes of Mn and Ga are not very
different, one expects small changes in the \hbox{Mn--O--Mn} angles,
and therefore, on the values of the couplings $t$ and
$J_\mathrm{AF}$. To keep the model simple we shall assume that the
parameter $J_\mathrm{AF}$ and $t$ are not modified at all by Ga
substitution.  Since the series LaGa$_y$Mn$_{1-y}$O$_3$ exist for all
$y$, we expect Ga to be in a Ga$^{3+}$ oxidation state, which is a
filled-shell configuration. This means that the corresponding
$\mathbf{e_g}$ orbital will not be available for a hole in a
neighboring Mn to hop, which reduces the Double-Exchange (DE)
ferromagnetic interaction. We thus encounter a quantum-percolation
problem (see e.g. Ref.~\onlinecite{QUANTUMPERCOLATION}). However,
quantum-percolation studies have shown that the quantum threshold is
not far from the classical one (that would be reached at 69\% Ga
fraction!), and this effect is not expected to be crucial. Moreover,
given the filled-shell configuration of Ga, the Mn$^{3+}$ core spin
$\mbox{\boldmath{$S$}}_i$ will disappear from Ga-sites, which reduces
also the AFM interaction that competes with the DE ferromagnetic
one. In the case of Ga, the effective AF coupling of Eq.~(\ref{jeff})
is $J_{\mathrm{eff}}=(1-y)^2J_{\mathrm{AF}}$ which, for instance, for
$y=0.12$ turns out to be 0.77 $J_{\mathrm{AF}}$. This is really a huge
reduction (see Fig.~3 of Ref.~\onlinecite{Alonso00a} and Fig.~3 of
Ref.~\onlinecite{Alonso01a}), and it is not obvious what the effect
of Ga substitution on the Curie temperature would
be. Fortunately, the effects of Ga substitution described so far can
be straightforwardly studied with the VMF method,\cite{Alonso00b} and
it turns out that the reduction on the antiferromagnetic interaction
has an stronger effect than the reduction of the DE mechanism, and
$T_{\mathrm C}$ {\em rises} in the model when the Ga fraction is
increased. Since this is in plain contradiction with
experiments,\cite{SUN99} it is clear that our microscopic description
of the Ga substitution is still incomplete.  This suggests that an
on-site electrostatic perturbation is induced by Ga ions in
Eq.~(\ref{hamil}).  This electrostatic term can be justified by the
fact that the average charge on Ga sites is always $+3$ while on Mn
sites it is a mixture of $+3$ and $+4$.  Therefore holes cannot hop
onto Ga sites, but they are attracted to them.

We expect this phenomenon to be more relevant in the case of Ga than
in the case of Fe. In fact, if in a cubic lattice of Mn ions we
substitute a fraction $y$ of Mn by TR randomly, the electrostatic
potential that appears in the Mn sites is proportional to the number
of TR$^{3+}$ that are neighbors of the considered Mn ion. 
Let us call $\epsilon$ the electrostatic potential felt by a hole at a Mn site
with only a neighboring TR. In the Fe case, $\epsilon$ is the same on
each Mn site because lattice distortion effects can be ignored, ought
to the similar ionic radii of Fe$^{3+}$ and Mn$^{3+}$. However, in the
Ga case (and even more in the Al case), the ionic radii difference
implies that to the randomness in the distribution is added the
randomness originated by the lattice distortion which, among other
phenomena, makes the potential randomly dependent on the Mn position.

To avoid an overly complicated model, we incorporate the second disorder
to the first one (of Lifshitz type) and we work with an effective
$\epsilon$, the same for all Mn sites. Consequently, the $\epsilon$ is
larger for Ga and Al than for Fe; this is precisely what the
experimental data require \cite{SUN99} since, as we have said, without
a localization mechanism that destroys the double exchange, $T_{\mathrm C}$
would rise with $y$. 

As we will see, a moderate disorder (from our analysis, $\epsilon<2t$) 
modifies only slightly the decrease of $T_{\mathrm C}$ with $y$. For this reason
we have neglected it in the case of Fe.

In summary, our model for 
La$_{2/3}$Ca$_{1/3}$Mn$_{1-y}$Ga$_y$O$_3$ is as follows:
every Ga atom modifies the Hamiltonian (\ref{hamil}) in the following ways:
\begin{enumerate}
\item Holes cannot hop onto Ga. Therefore the average number of one
third of a hole per unit cell, actually means that the fraction of
Mn$^{4+}$ is increased with respect to the case $y=0$ by a factor
\hbox{$(1-y)^{-1}$}.
\item The core spins vanish at the Ga sites.
\item A electrostatic potential appears on the Mn sites,
proportional to the number of Ga$^{3+}$ that are nearest-neighbors of
the considered Mn site. 
\end{enumerate}

\section{Results \label{sec:results}}

We have studied our disordered model using the VMF
technique,\cite{Alonso00b} which is approximate for the spins, but
treat charge carriers on each spin texture exactly (up to controlled
numerical errors).  The Hybrid Monte Carlo study~\cite{Alonso00c} has
shown that the VMF overestimates the critical temperature by a 30\%,
which is the same overestimation factor found in classical
statistical mechanics three-dimensional models with short range
interactions. In the particular case of a disordered model, the VMF
technique has the important merit of allowing the study of very large
clusters ($96\times 96\times 96$ in this work). The self-averaging
nature of the electronic density of states makes thus unnecessary to
average our results over disorder realizations, which instead
would be mandatory in a Monte Carlo study where the study of a single
$16\times 16\times 16$ cluster is at the very limit of present day
computers and algorithms.\cite{Alonso00c} In order to make the study
as unbiased as possible, we shall consider as VMF ansatzs all the spin
orderings that were found in the Monte Carlo study of model
in the absence of disorder,\cite{Alonso01a} namely FM, G-AFM, A-AFM,
C-AFM, skyrmion (SK), flux, twisted and island phases.

\begin{figure}
\includegraphics[angle=90,width=\columnwidth]{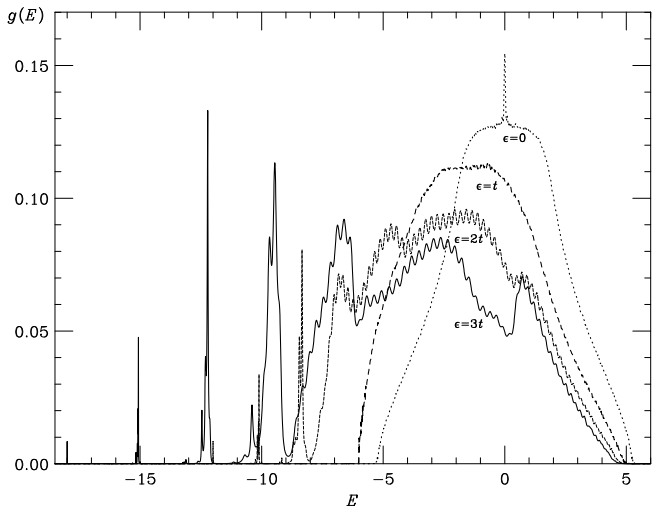}
\caption{Hole density of states for a full spin-polarized
system with a 20\% Gallium ($y=0.2$), and several values of the
parameter $\epsilon$.  The curves are normalized to a maximum of $1-y$ holes
per unit cell, and the Fermi level is obtained integrating the DOS
until reaching one third of a hole per unit cell.  }
\label{FIG1}
\end{figure}

To gain some insight on the physics of our model, it is useful to
consider (see Fig.~\ref{FIG1}) the hole density of states
(DOS)\cite{Benoit92,convergence} for the fully FM spin-polarized
system with a 20\% Ga substitution, upon variation of the
electrostatic parameter $\epsilon$.  For $\epsilon=0$, we have the
usual DOS of the cubic lattice, slightly contracted and smoothed by
the percolative disorder (holes cannot jump onto a Ga). We have a
sharp contribution at midband, due to the localized states produced by
the rare configurations of Mn sites fully surrounded by Ga sites
(thereafter this very rare configurations were taken out by hand).  If
the electrostatic potential at a given site is large enough (in a Mn
site with $k$ neighboring Ga, it is $k\epsilon$), a hole state of
energy $-k\epsilon$ will get localized in this Mn site, as the reader
can check in Fig.~\ref{FIG1}. This effect can be investigated within
the small defect-density expansion,\cite{Martin-Mayor00} where $\epsilon$
is allowed to be arbitrarily large. It turns out that a localized
state will form if $k\epsilon$ is larger than the edge of the
$\epsilon=0$ DOS.\cite{Martin-Mayor00} Therefore, Fig.~\ref{FIG1}
tells us that for $\epsilon=2t$, a Mn site should have at least three
neighboring Ga to trap a hole on it, while for $\epsilon=3t$ it is
enough to have two neighboring Ga (Mn sites with two neighboring Ga
sites are deceivingly common, even for modest values of $y$, because
of the large combinatorial factor).  The electrostatic potential tends
also to localize the electronic states even near to the edges of the
main part of the DOS, depicted in Fig.~\ref{FIG1} (for $y=0.2$ the
Fermi level is very close to the edge of the main part of the DOS).
The calculations show that the double exchange interaction is strongly
suppressed when holes occupy these localized states, leading to a
decrease in the value of the $T_{\mathrm C}$. We find that the drop in
$T_{\mathrm C}$ upon Ga doping is best reproduced using $\epsilon \sim
3 t$, and, considering our earlier results for manganites with a
perfect Mn lattice,\cite{Alonso00a,Alonso00b,Alonso00c,Alonso01a} we
take $J_{\mathrm{AF}} \sim 0.06t$---$0.07t$.  On the other hand, for
$\epsilon<2 t$ disorder is not the main agent of the eventual decrease
of $T_{\mathrm C}$ as $y$ increases (as happens in the case
TR\,=\,Fe).

\begin{figure}
\includegraphics[angle=90,width=\columnwidth]{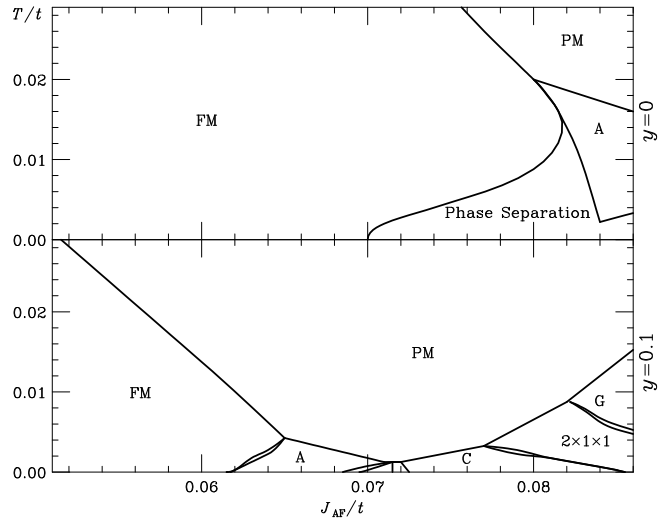}
\caption{Magnetic phase-diagram for our model at $x=1/3$, as function
of $J_\mathrm{AF}$, for two values of the Ga concentration, $y$. Top:
$y=0$. Bottom: $y=0.1$.}
\label{FIG2}
\end{figure}

In Fig.~\ref{FIG2} we show the phase diagram of model (\ref{hamil}),
for 1/3 hole concentration, in the cases of $y=0$ (top) and $y=0.1$
(bottom) Mn---Ga substitution. The very narrow phase-separation
regions at $y=0.1$ are left unlabelled for clarity. In agreement with
the expected role of disorder near a discontinuous transition, Ga
substitution induces a wedge of PM phase, down to the lowest
temperatures, between phases with different types of long range
magnetic order. In the $0.06t$---$0.07t$ $J_{\mathrm{AF}}$ range the
SK and A-AFM phases are virtually degenerate. Between the A and C
phases we find tiny regions of phase-separation, $2\times 2\times 2$
island phase and twisted orderings. Notice that the VMF assumes that
the spin structure is spatially homogeneous, overestimating the
tendency towards first order transitions in the disordered
system. Thus, the results shown in Fig.~\ref{FIG2} strongly suggest
(see below) the existence of a continuous transition at all
temperatures, in the Ga doped system.

\begin{figure}
\includegraphics[angle=90,width=\columnwidth]{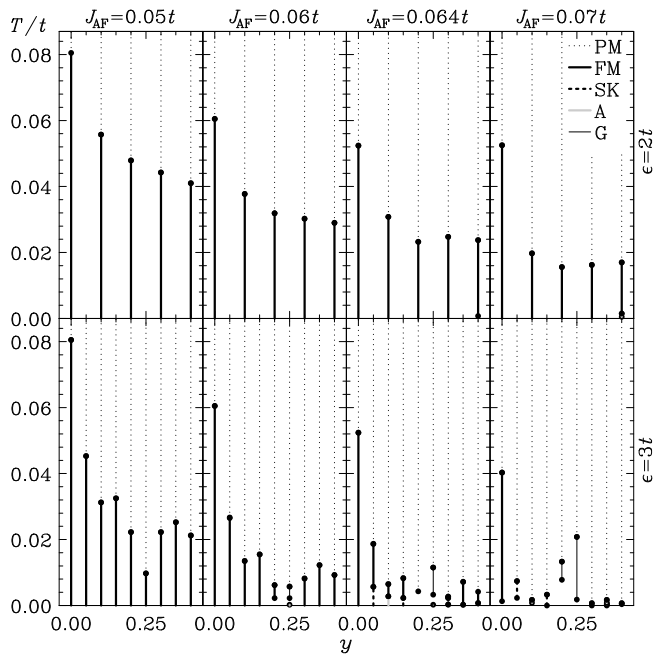}
\caption{ Bottom: phase-diagram for several values of $J_\mathrm{AF}$,
including $J_\mathrm{AF}=0.05t$
(La$_{2/3}$Sr$_{1/3}$Mn$_{1-y}$Ga$_y$O$_3$) and
$J_\mathrm{AF}=0.064t$---$0.07t$
(La$_{2/3}$Ca$_{1/3}$Mn$_{1-y}$Ga$_y$O$_3$), as predicted by our
model. Top: phase-diagram for electrostatic potential $\epsilon =
2 t$.}
\label{FIG3}
\end{figure}

In figure~\ref{FIG3} (bottom) we show the phase diagram of our model,
for four values of $J_\mathrm{AF}$, namely $J_\mathrm{AF}=0.05t$,
which is a reasonable value for
La$_{2/3}$Sr$_{1/3}$Mn$_{1-y}$Ga$_y$O$_3$ at least for
$y=0$,\cite{Alonso00a} $J_\mathrm{AF}=0.06t$ that marks the change
from second to first order character of the PM-FM transition, and
$J_\mathrm{AF}=0.064t$---$0.07t$ which we believe is an appropriate
range for La$_{2/3}$Ca$_{1/3}$Mn$_{1-y}$Ga$_y$O$_3$. There are several
points to be noted. First, see how at $y=0$, the different values of
$J_\mathrm{AF}$ are enough to explain the 30\% differences in
$T_{\mathrm C}$ between La$_{2/3}$Sr$_{1/3}$MnO$_3$ and
La$_{2/3}$Ca$_{1/3}$MnO$_3$.  As we see, for
La$_{2/3}$Sr$_{1/3}$Mn$_{1-y}$Ga$_y$O$_3$, $J_{\mathrm{AF}} \sim
0.05t$, we have a more or less monotonous decreases of the $T_{\mathrm
C}$, the ground-state being FM in all the interesting $y$ range. On
the other hand, the scenario is completely different for
La$_{2/3}$Ca$_{1/3}$Mn$_{1-y}$Ga$_y$O$_3$ ($J_\mathrm{AF}=0.064t$ and
$0.07t$). In this case, 
the FM ordering disappears for $y_{\mathrm{c}}\approx 0.1$---$0.2$,
where the PM phase is stable until the lowest temperatures.
{\em For larger values of $y$ a very complicated situation is reached
with different, almost degenerated phases}. This region might look
somehow glassy experimentally. The ordering temperature in the large
$y$ region is about one order of magnitude smaller than for
$y=0$. More remarkably, for $y$ larger than $y_{\mathrm{c}}$, the
spins are not completely polarized even at zero temperature
(technically, this means that the mean-field at low temperature is
$\alpha T$, where $\alpha$ is a large but not infinite
constant~\cite{Alonso00b}), and, also at zero temperature, the
magnetization in the FM phase decrease upon approaching
$y_{\mathrm{c}}$ although it does not get below $80\%$
polarization. This effect will be amplified by the neglected quantum
fluctuations of the spins, especially at very low temperatures. Thus,
{\em given the wedge of PM phase that reaches near zero temperature at
$y_{\mathrm{c}}$ and our underestimation of quantum fluctuations, it
seems fairly plausible that a quantum critical point will be found in}
La$_{2/3}$Ca$_{1/3}$Mn$_{1-y}$Ga$_y$O$_3${\em, for $y=0.1$---$0.2$}.
In order to emphasize how crucial the electrostatic attraction of
holes to the Ga sites is, let us show in Fig.~\ref{FIG3} (top), the
equivalent of Fig.~\ref{FIG3} (bottom), for $\epsilon=2t$. In this
case, the system is ferromagnetic in all the explored range $0\le y
\le 0.4$, in agreement with the scarcity of localized states shown in
Fig.~\ref{FIG1}.

\section{Remarks and predictions \label{sec:remarks}}

In this work we have accomplished a systematic study of doped
manganites begun in previous papers and formulated a microscopically
motivated disordered model to study the effects of Ga substitution in
La$_{2/3}$Ca$_{1/3}$Mn$_{1-y}$Ga$_y$O$_3$.  We have studied this model
using the VMF technique in $96\times 96\times 96$ clusters. A result
of this work is the prediction that a QCP is expected to appear for
$y_{\mathrm{c}}\approx 0.1$---$0.2$, which is induced by Anderson
localization of the electronic states below the Fermi level that
suppresses the double exchange mechanism. Thus, we expect localization
effects in the transport properties at low temperatures, where phonons
are static ({\em i.e} polarons could exist~\cite{Verges01}, but they
would play the same role of our $\epsilon$ diagonal static disorder
term).  For larger values of $y$, a different phase is reached which
is difficult to analyze because many different phases are almost
degenerate. Nevertheless one can conclude safely that the ordering
temperature for $0.3<y<0.4$ will be in the 20 to 40 K range. It is
worth noting that the main features of the predicted phase-diagram are
in quantitative agreement with preliminary experimental
work.\cite{Algarabel02}

A completely different behavior is expected for
La$_{2/3}$Sr$_{1/3}$Mn$_{1-y}$Ga$_y$O$_3$. In such case the
localization phenomenon that can be inferred from Fig.~\ref{FIG1} for
$\epsilon=3t$, besides to the results of Fig.~\ref{FIG3} corresponding
to $J_{\mathrm{AF}}=0.05t$, strongly suggest that the fact that holes
are localized does not necessarily imply that the long range magnetism
disappears; in fact, to maintain it, it is enough that effective
ferromagnetic interactions between close spins remain. Hence, for
La$_{2/3}$Sr$_{1/3}$Mn$_{1-y}$Ga$_y$O$_3$ we expect a
FMmetal-FMinsulator transition at low temperatures (see also
Ref.~\onlinecite{Verges01}).

In the Ga case, it is worthwhile to remark that in our model, which 
considers spins coupled to charge carriers, the localization transition
determines the magnetic critical exponents too. Experimentally, it is
very promising that one be able to study a localization transition
through a magnetic order parameter. If there is really a QCP for 
$y\approx 0.1$---$0.2$,\cite{Algarabel02} there would be presumably only one
characteristic length in the system, which would set the scale both of
the magnetic and transport phenomena.

It is therefore interesting the theoretical and experimental
determination of the magnetic critical exponents associated to the
magnetic transition at very low $T$ and its universality class.

Finally, it is worth noting that the model used in this work includes
the simplest combination of interactions compatible with the 
experimental data for the manganites, in the absence of Ga doping. 
The confirmation of the predictions proposed here for Ga doped materials
can illustrate the usefulness of simple models in studying
the phase diagram of these compounds.

We thank P. A. Algarabel, J. Blasco, J. Garc\'{\i}a, R. Ibarra,
J. M. de Teresa and for discussions.  The authors thank financial
support from CICyT (Spain) through grants PB96-0875, AEN99-0990,
FPA2000-0956 and FPA2000-1252.  V.M.-M. was supported by E.C. contract
HPMF-CT-2000-00450.


\begin{thebibliography}{99}
\bibitem{phonons}
Phonons are crucial to understand magnetoresistance
effects, see  Ref.~\onlinecite{Verges01} and references
therein.
\bibitem{convergence} 
The presence of gaps in the DOS makes the convergence of the
continued-fraction~\cite{Benoit92} rather slow, which is the reason
for the small oscillations shown in Fig~\ref{FIG1}. Nevertheless
integrated quantities, and in particular the location of phase
transitions, are very slightly affected.
\end{thebibliography}
\end{document}